\documentclass[11pt]{article}

\usepackage{fullpage,amsmath,amsfonts,latexsym}

\def\bra#1{\left<#1\right|}
\def\ket#1{\left|#1\right>}
\def\braket#1#2{\left<#1|#2\right>}
\def\ketbra#1#2{\left|#1\right>\left<#2\right|}
\newtheorem{theorem}{Theorem}[section]
\newtheorem{corollary}[theorem]{Corollary}

\newtheorem{definition}[theorem]{{\bf Definition}}

\newenvironment{proof}
{\noindent {\bf Proof }}
{{\hfill $\Box$}\\
 \smallskip}

\def\S{{\cal S}}
\def\PQC{{\bf PQC}}
\def\H{{\cal H}}
\def\rn{{\rho_0}}
\def\ra{{\rho_a}} 
\def\E{{\cal E}}
\def\C{{\cal C}}

\def\Tr{{\rm Tr}}
\def\01{\{0,1\}}
\def\Hp#1{H(p_1,\ldots,p_{#1})}
\def\Hq#1{H(q_1,\ldots,q_{#1})}

\title{Private Quantum Channels  \\
and the Cost of Randomizing Quantum Information}
\author{
Michele Mosca\thanks{University of Waterloo, {\tt mmosca@cacr.math.uwaterloo.ca}.
Partially supported by NSERC.}
\and 
Alain Tapp\thanks{University of Waterloo, {\tt atapp@cacr.math.uwaterloo.ca}.
Supported by NSERC Postdoctoral Fellowship.}
\and 
Ronald de Wolf\thanks{CWI and University of Amsterdam, {\tt rdewolf@cwi.nl}. 
Partially supported by the EU-project QAIP.}
}
\date{\today}
\begin{document}

\maketitle

\begin{abstract}
We investigate how a classical private key can be used by two players, 
connected by an insecure one-way quantum channel, to perform private 
communication of quantum information. 
In particular we show that in order to transmit $n$ qubits privately, 
$2n$ bits of shared private key are necessary and sufficient.
This result may be viewed as the quantum analogue of the classical 
one-time pad encryption scheme. From the point of view of the eavesdropper, 
this encryption process can be seen as a randomization of the original state.
We thus also obtain strict bounds on the amount of entropy necessary 
for randomizing $n$ qubits.

\end{abstract}

\section{Introduction}

Secure transmission of classical information is a well studied topic.
Suppose Alice wants to send an $n$-bit message $M$ to Bob over an insecure
(i.e.~spied-on) channel, in such a way that the eavesdropper Eve cannot 
obtain any information about $M$ from tapping the channel.
If Alice and Bob share some secret $n$-bit key $K$, then here is
a simple way for them to achieve their goal:
Alice exclusive-ors $M$ with $K$ and sends the result $M'=M\oplus K$ 
over the channel, Bob then xors $M'$ again with $K$ and obtains 
the original message $M'\oplus K=M$.
Eve may see the encoded message $M'$, but if she does not know $K$
then this will give her no information about the real message $M$,
since for any $M$ there is a key $K'$ giving rise to the same encoding $M'$.
This scheme is known as the {\em Vernam cipher} or {\em one-time pad}
(``one-time'' because $K$ can be used only once if we want 
information-theoretic security).
It shows that $n$ bits of shared secret key are sufficient to securely transmit 
$n$ bits of information. 
Shannon~\cite{shannon:communication,shannon:secrecy} 
has shown that this scheme is optimal:
$n$ bits of shared key are also {\em necessary} in order to transmit
an $n$-bit message in an information-theoretically secure way.

Now let us consider the analogous situation in the quantum world.
Alice and Bob are connected by a one-way quantum channel, 
to which an eavesdropper Eve has complete access.
Alice wants to transmit to Bob some $n$-qubit state $\rho$ taken 
from some set $\S$, without allowing Eve to obtain any information about $\rho$.
Alice and Bob could easily achieve such security if they share $n$ EPR-pairs
(or if they were able to establish EPR-pairs over a secure quantum channel),
for then they can apply teleportation~\cite{teleporting} and transmit every
qubit via 2 random classical bits, which will give Eve no information whatsoever.
But now suppose Alice and Bob do not share EPR-pairs, but instead they only have
the resource of shared randomness, which is weaker but easier to maintain.

A first question is: is it at all possible to send quantum information
fully securely using only a finite amount of randomness?
At first sight this may seem hard: Alice and Bob have to ``hide'' the
amplitudes of a quantum state, which are infinitely precise complex numbers.
Nevertheless, the question has a positive answer.
More precisely, to privately send $n$ qubits, a $2n$-bit classical 
key is sufficient. The encryption technique is fairly natural. 
Alice applies to the state $\rho$ she wants to transmit a reversible quantum 
operation specified by the shared key $K$ (basically, she applies a random
Pauli matrix to each qubit), and she sends the result $\rho'$ to Bob.
In the most general setting this reversible operation can be represented as
doing a unitary operation on the state $\rho$ augmented with a known fixed
ancilla state $\ra$.
Knowing the key $K$ that Alice used, Bob knows which operation Alice applied
and he can reverse this, remove the ancilla, and retrieve $\rho$.
In order for this scheme to be information-theoretically secure against 
the eavesdropper, we have to require that Eve always ``sees'' the same 
density matrix $\rn$ on the channel, no matter what $\rho$ was.
Because Eve does not know $K$, this condition can indeed be satisfied.
Accordingly, an insecure quantum channel can be made secure (private)
by means of shared classical randomness.

A second question is, then, {\em how much} key Alice and Bob need to share in
order to be able to privately transmit any $n$-qubit state.
A good way to measure key size is by the amount of entropy required to create it.
As one might imagine, showing that $2n$ bits of key are also necessary is the most 
challenging part of the article. We prove this in Section~\ref{secPQClowerbound}.%
\footnote{\label{noteQKD}Note that if Alice and Bob share an insecure {\em
two}-way channel, then they can do quantum key exchange~\cite{bb84} in order 
to establish a shared random key, so in this case no prior shared key 
(or only a very small one) is required.}
Accordingly, in analogy with the classical one-time pad, we have an optimal quantum
one-time pad which uses $2n$ classical bits to completely ``hide'' $n$ qubits from Eve.
In particular, hiding a qubit is only twice as hard as hiding a classical bit, 
despite the fact that in the qubit we now have to hide amplitudes coming 
from a continuous set.

Now imagine an alternative scenario. 
Alice has a state $\rho$ from some specific set and she wants to randomize it completely.
How much entropy does she need for this?
That is, what is the thermodynamical cost of forgetting quantum information?
A natural and general way to do that is for Alice to perform a unitary transformation 
to $\rho$ augmented with an ancilla and then to forget which one.
The thermodynamical price of this operation is now the entropy of the
probability distribution over the set of unitary transformations. 
The parallel between these two scenarios should be clear.
If one has a private quantum channel, 
one automatically has a related randomization procedure.
Consequently, we obtain the result that $2n$ bits are necessary
and sufficient to randomize an $n$-qubit quantum register. 
For the case $n=1$, this result has also been obtained by 
Braunstein, Lo, and Spiller~\cite{bls:forget,lo:generalization}.

The article is organized as follows.
Section~\ref{vne} introduces some notation and some properties of Von Neumann entropy.
In Section~\ref{mains} we give a formal definition of a private quantum channel (\PQC).
In Section~\ref{secPQCupperbound} we give some examples of \PQC s.
In particular we show that there is a \PQC\ that privately sends 
any $n$-qubit state using $2n$ bits of randomness (shared key).
We also exhibit a non-trivial set of $n$-qubit states (namely
the tensor products of qubits with real amplitudes) for which there
is \PQC\ requiring only $n$ bits of randomness.
The latter result includes the classical one-time pad.
In Section~\ref{secPQClowerbound} we show that $2n$ bits of randomness 
are necessary if we want to be able to send any $n$-qubit state privately.
Finally, in Section~\ref{rnd} we restate the previous results in terms of 
the thermodynamical cost of randomization of quantum information.

\bigskip

{\bf Remark about related work.}
A number of recent papers independently discussed issues similar to our work.
We already mentioned the result of Braunstein, Lo, and 
Spiller~\cite{bls:forget,lo:generalization} for state randomization.
Very recently, Boykin and Roychowdhury~\cite{boykin&roychowdhury:encryption}
exhibited the $2n$-bit Pauli-matrix one-time pad and proved a $2n$-bit lower 
bound for the case where the encryption scheme does not allow the use of 
an ancilla state (they also give a general characterization of all possible 
encryption schemes without ancilla).
In Section~\ref{secPQClowerbound} we give a simpler proof of this lower
bound for the no-ancilla case and give a different and more complicated
proof for the lower bound in the case where we do allow an ancilla.

\section{Preliminaries}\label{vne}

\subsection{States and operators}\label{ssecstoper}

We use $||v||$ for the Euclidean norm of vector $v$.
If $A$ is a matrix, then we use $A^{\dagger}$ for its conjugate transpose
and $\Tr(A)$ for its trace (the sum of its diagonal entries).
A square matrix $A$ is {\em Hermitian} if $A=A^{\dagger}$, 
and {\em unitary} if $A^{-1}=A^{\dagger}$.
Important examples of unitary transformations are the 4 {\em Pauli matrices}:
$$
\sigma_0=\left(\begin{array}{cc}1 & 0\\ 0 & 1\end{array}\right), \
\sigma_1=\left(\begin{array}{cc}0 & 1\\ 1 & 0\end{array}\right), \
\sigma_2=\left(\begin{array}{cc}0 & -i\\ i & 0\end{array}\right), \
\sigma_3=\left(\begin{array}{cc}1 & 0\\ 0 & -1\end{array}\right).
$$ 
Let $\ket{0},\ldots,\ket{M-1}$ denote the basis states of some $M$-dimensional
Hilbert space $\H_M$. We use $\H_{2^n}$ for the Hilbert space whose
basis states are the $2^n$ classical $n$-bit strings.
A {\em pure quantum state} $\ket{\phi}$ is a norm-1 vector in $\H_M$. 
We treat $\ket{\phi}$ as an $M$-dimensional column vector and
use $\bra{\phi}$ for the row vector that is its conjugate transpose.
The {\em inner product} between pure states $\ket{\phi}$ and $\ket{\psi}$
is $\braket{\phi}{\psi}$.
A {\em mixed quantum state} or {\em density matrix} $\rho$ 
is a non-negative Hermitian matrix that has trace $\Tr(\rho)=1$.
The density matrix corresponding to a pure state $\ket{\phi}$ is
$\ketbra{\phi}{\phi}$.
Because a density matrix $\rho$ is Hermitian, it has a diagonalization
$\rho=\sum_{i=1}^{N} p_i\ketbra{\phi_i}{\phi_i}$, 
where the $p_i$ are its eigenvalues, $p_i\geq 0$, $\sum_i p_i=1$, 
and the $\ket{\phi_i}$ form an orthonormal set.
Thus $\rho$ can be viewed as describing a probability distribution over pure states.
We use $\tilde{I}_{M}=\frac{1}{M}I_{M}=\frac{1}{M}\sum_{i=1}^M\ketbra{i}{i}$ 
to denote the totally mixed state, which represents the uniform distribution on
all basis states.
If two systems are in pure states $\ket{\phi}$ and $\ket{\psi}$, respectively,
then their joint state is the tensor product pure state 
$\ket{\phi}\otimes\ket{\psi}=\ket{\phi}\ket{\psi}$.
If two systems are in mixed states $\rho_1$ and $\rho_2$, respectively,
then their joint state is the tensor product $\rho_1\otimes\rho_2$.
Note that $(\ket{\phi} \otimes \ket{\psi})(\bra{\phi}\otimes\bra{\psi})$
is the same as $\ketbra{\phi}{\phi}\otimes\ketbra{\psi}{\psi}$.

Applying a unitary transformation $U$ to a pure state $\ket{\phi}$ gives 
pure state $U\ket{\phi}$, applying $U$ to a mixed state $\rho$ 
gives mixed state $U\rho U^{\dagger}$.
We will use $\E=\{\sqrt{p_i}U_i\mid 1\leq i\leq N\}$ to denote 
the {\em superoperator} which applies $U_i$ with probability $p_i$ to 
its argument (we assume $\sum_i p_i=1$).
Thus $\E(\rho)=\sum_i p_i U_i\rho U_i^\dagger$.
Quantum mechanics allows for more general superoperators, 
but this type suffices for our purposes.
If two superoperators 
$\E=\{\sqrt{p_i}U_i\mid 1\leq i\leq N\}$ and
$\E'=\{\sqrt{p'_i}U'_i\mid 1\leq i\leq N'\}$ 
are identical ($\E(\rho)=\E'(\rho)$ for all $\rho$), then they are
unitarily related in the following way~\cite[Section~3.2]{nielsen:thesis}
(where we assume $N\geq N'$ and if $N>N'$ we pad $\E'$ with 
zero operators to make $\E$ and $\E'$ of equal size):
there exists a unitary $N\times N$ matrix $A$ such that for all $i$
$$
\sqrt{p_i}U_i = \sum_{j=1}^N A_{ij} \sqrt{p'_j}U'_j.
$$

\subsection{Von Neumann entropy}

Let density matrix $\rho$ have the diagonalization 
$\sum_{i=1}^{N}p_i\ketbra{\phi_i}{\phi_i}$. 
The {\em Von Neumann entropy} of $\rho$
is $S(\rho)=\Hp{N}=-\sum_{i=1}^N p_i \log p_i$, where $H$ is the classical entropy function.
This $S(\rho)$ can be interpreted as the minimal Shannon entropy of 
the measurement outcome, minimized over all possible complete measurements.
Note that $S(\rho)$ only depends on the eigenvalues of $\rho$.
The following properties of Von Neumann entropy will be useful later
(for proofs see for instance~\cite{Wehrl:entropy}).
\begin{enumerate}
\item $S(\ketbra{\phi}{\phi})=0$, for every pure state $\ket{\phi}$.
\item $S(\rho_1\otimes\rho_2)=S(\rho_1)+S(\rho_2)$.
\item $S(U \rho \, U^{\dagger})=S(\rho)$.
\item $S(\lambda_1 \rho_1 + \lambda_2 \rho_2 + \cdots + \lambda_n \rho_n) \geq 
       \lambda_1 S(\rho_1) + \lambda_2 S(\rho_2) + \cdots + \lambda_n S(\rho_n)$
if $\lambda_i\geq 0$ and $\sum_i\lambda_i=1$.
\item If $\rho = \sum_{i=1}^{N} p_i \, \ketbra{\phi_i}{\phi_i}$ with
the $\ket{\phi_i}$ not necessarily orthogonal, then $S(\rho)\leq \Hp{N}$.
\end{enumerate}

\section{Private Quantum Channel}\label{mains}

Let us sketch the scenario for a private quantum channel.
There are $N$ possible keys, which we identify for convenience with 
the numbers $1,\ldots,N$. The $i$th key has probability $p_i$, 
so the key has entropy $\Hp{N}$ when viewed as a random variable.
Each key $i$ corresponds to a unitary transformation $U_i$.
Suppose Alice wants to send a pure state $\ket{\phi}$ from some set $\S$ to Bob.
She appends some fixed ancilla qubits in state $\ra$ to $\ketbra{\phi}{\phi}$
and then applies $U_i$ to $\ketbra{\phi}{\phi}\otimes\ra$, where $i$ is her key.
She sends the resulting state to Bob.
Bob, who shares the key $i$ with Alice, applies $U_i^{-1}$ 
to obtain $\ketbra{\phi}{\phi}\otimes\ra$, removes 
the ancilla $\ra$, and is left with Alice's message $\ketbra{\phi}{\phi}$.
Now in order for this to be secure against an eavesdropper Eve,
we have to require that if Eve does not know $i$, then the density matrix $\rn$
that she gets from monitoring the channel is independent of $\ket{\phi}$.
This implies that she gets no information at all about $\ket{\phi}$.
Of course, Eve's measuring the channel might destroy the encoded message,
but this is like classically jamming the channel and cannot be avoided.
The point is that {\em if\/} Eve measures, then she receives no information
about $\ket{\phi}$.
It is not hard to see that this is the most general quantum mechanical scenario 
which allows Bob to recover the message perfectly and at the same time gives 
Eve zero information.

We formalize this scenario as follows.

\begin{definition}
Let $\S \subseteq \H_{2^n}$ be a set of pure $n$-qubit states, 
$\E=\{ \sqrt{p_i} U_i \mid  1\leq i \leq N \}$ be a superoperator
where each $U_i$ is a unitary mapping on $\H_{2^m}$, $\sum_{i=1}^N p_i=1$, 
$\ra$ be an $(m-n)$-qubit density matrix, and $\rn$ be an $m$-qubit density matrix. 
Then $[\S,\E,\ra,\rn]$ is called a {\em Private Quantum Channel (\PQC)} 
if and only if for all $\ket{\phi}\in\S$ we have
$$
\E(\ketbra{\phi}{\phi} \otimes \ra)=
\sum_{i=1}^N p_i U_i \left( \ketbra{\phi}{\phi} \otimes \ra \right)
U_i^\dagger = \rn.
$$
If $n=m$ (i.e.~no ancilla), then we omit $\ra$.
\end{definition}

Note that by linearity, if the \PQC\ works for all pure states in $\S$, 
then it also works for density matrices over $\S$: applying the \PQC\ to a
mixture of states from $\S$ gives the same $\rn$ as when we apply it to a pure state.
Accordingly, if $[\S,\{ \sqrt{p_i} \, U_i \mid  1\leq i \leq N \},\ra,\rn]$ 
is a \PQC, then $\Hp{N}$ bits of shared randomness are sufficient for Alice 
to send any mixture $\rho$ of $\S$-states to Bob in a secure way.
Alice encodes $\rho$ in a reversible way depending on her key $i$ and Bob 
can decode because he knows the same $i$ and hence can reverse Alice's
operation $U_i$.
On the other hand, Eve has no information about the key $i$ apart from the
distribution $p_i$, so from her point of view the channel is in state $\rho_{Eve} = \rn$.
This is independent of the $\rho$ that Alice wants to send,
and hence gives Eve no information about $\rho$.

\section{Examples of Private Quantum Channels}\label{secPQCupperbound}

In this section we exhibit some private quantum channels.
The first uses $2n$ bits key to send privately any $n$-qubit state.
The idea is simply to apply a random Pauli matrix to each bit individually.
This takes 2 random bits per qubit and 
it is well known that the resulting qubit is in the completely mixed state.
For notational convenience we identity the numbers 
$\{0,\ldots,2^{2n}-1\}$ with the set $\{0,1,2,3\}^n$.
For $x\in\{0,1,2,3\}^n$ we use $x_i\in\{0,1,2,3\}$ for its $i$th entry,
and we use $\overline{\sigma_x}$ to denote the $n$-qubit unitary transformation
 $\sigma_{x_1}\otimes\cdots\otimes\sigma_{x_n}$.

\begin{theorem}\label{sufficient}
If $\E=\{\frac{1}{\sqrt{2^{2n}}}\overline{\sigma_x} \mid x\in\{0,1,2,3\}^n\}$, 
then $[\H_{2^n},\E,\tilde{I}_{2^n}]$ is a \PQC.
\end{theorem}

\begin{proof}
It is easily verified that applying each $\sigma_i$ with probability $1/4$
to a qubit puts that qubit in the totally mixed state $\tilde{I}_2$
(no matter if it is entangled with other qubits).
Operator $\E$ just applies this treatment to each of the $n$ qubits, hence
$\E(\ketbra{\phi}{\phi})=\tilde{I}_{2^n}$ for every $\ket{\phi}\in\H_{2^n}$.
\end{proof}

Since the above $\E$ contains $2^{2n}$ operations and they have uniform probability, it 
follows that $2n$ bits of private key suffice to privately send any state from $\H_{2^n}$. 

The next theorem shows that there is some nontrivial 
subspace of $\H_{2^n}$ where $n$ bits of private key suffice,
namely the set of all tensor products of real-amplitude qubits.

\begin{theorem}\label{reel}
If $B=\{\cos(\theta)\ket{0}+\sin(\theta)\ket{1}\mid 0\leq\theta<2\pi\}$,
$\S=B^{\otimes n}$, and $\E=\{\frac{1}{\sqrt{2^n}}\overline{\sigma_x}\mid x\in\{0,2\}^n\}$,
then $[\S,\E,\tilde{I}_{2^n}]$ is a \PQC.
\end{theorem}

\begin{proof}
This is easily verified: applying $\sigma_0$ and $\sigma_2$, each
with probability 1/2, puts any qubit from $B$ in the totally mixed state.
Operator $\E$ does this to each of the $n$ qubits individually.
\end{proof}

Note that if we restrict $B$ to classical bits (i.e.~$\theta\in\{0,\pi/2\}$)
then the above \PQC\ reduces to the classical one-time pad: flipping each bit 
with probability 1/2 gives information-theoretical security.

In the previous \PQC s, $\rn$ was the completely mixed state $\tilde{I}_{2^n}$.  
This is no accident, and holds whenever $n=m$ and $\tilde{I}_{2^n}$ 
is one of the states that the \PQC\ can send:

\begin{theorem}\label{Mike}
If $[\S,\E,\rn]$ is a \PQC\ without ancilla and $\tilde{I}_{2^n}$ can be written 
as a mixture of $S$-states, then $\rn=\tilde{I}_{2^n}$.
\end{theorem}

\begin{proof}
If $\tilde{I}_{2^n}$ can be written as a mixture of $S$-states, then\\
\hspace*{15ex}
$\displaystyle
\rn=\E(\tilde{I}_{2^n})=
\sum_{i=1}^N p_iU_i\tilde{I}_{2^n} U_i^{\dagger}=
\sum_{i=1}^N \frac{p_i}{2^n}U_iU_i^{\dagger}=
\sum_{i=1}^N \frac{p_i}{2^n}I_{2^n}=\tilde{I}_{2^n}.$
\end{proof}

\noindent
In general $\rn$ need not be $\tilde{I}_{2^n}$.
For instance, let $\S=\{\ket{0},\frac{1}{\sqrt{2}}(\ket{0}+\ket{1})\}$, 
$\E=\{\sqrt{p_1}I_2, \frac{\sqrt{p_2}}{\sqrt{2}}\left(
\begin{matrix}
1 & 1 \\
1 & -1
\end{matrix}
\right)\}$ 
with $p_1=p_2=1/2$, and 
$\rn=\left(
\begin{matrix}
\frac{3}{4} & \frac{1}{4} \\
\frac{1}{4} & \frac{1}{4}
\end{matrix}
\right)$.
Then it is easily verified that $[\S,\E,\rn]$ is a \PQC.

\section{Lower Bound on the Entropy of \PQC s}\label{secPQClowerbound}

In the previous section we showed that $2n$ bits of entropy suffice
for a \PQC\ that can send arbitrary $n$-qubit states.
In this section we will show that $2n$ bits are also {\em necessary} for this. 
Very recently and independently of our work, this $2n$-bit lower bound was 
also proven by Boykin and Roychowdhury~\cite{boykin&roychowdhury:encryption}
for the special case where the \PQC\ is not allowed to use any ancilla qubits. 
We will first give a shorter version of their proof, 
basically by observing that a large part of it can be replaced
by a reference to the unitary equivalence of identical superoperators 
stated at the end of Section~\ref{ssecstoper}. 

\begin{theorem}\label{thnoancilla}
If $[\H_{2^n},\{ \sqrt{p_i} U_i \mid  1\leq i \leq N \}, \tilde{I}_{2^n}]$ 
is a \PQC, then $\Hp{N}\geq 2n$.
\end{theorem}

\begin{proof}
Let $\E=\{\sqrt{p_i}U_i\}$, 
$\E'=\{\frac{1}{\sqrt{2^{2n}}}\overline{\sigma_x} \mid x\in\{0,1,2,3\}^n\}$
be the superoperator of Theorem~\ref{sufficient}, and let $K=\max(2^{2n},N)$.
Since $\E(\rho)=\E'(\rho)=\tilde{I}_{2^n}$ for all $n$-qubit states $\rho$, 
we have that $\E$ and $\E'$ are unitarily related in the way mentioned
in Section~\ref{ssecstoper}: there exists a unitary $K\times K$ matrix $A$
such that for all $1\leq i\leq N$ we have
$$
\sqrt{p_i}U_i=\sum_{x\in\{0,1,2,3\}^n} A_{ix}\frac{1}{\sqrt{2^{2n}}}\overline{\sigma_x}.
$$
We can view the set of all $2^n\times 2^n$ matrices as a $2^{2n}$-dimensional
vector space, with inner product $\langle M,M'\rangle=\Tr(M^\dagger M')/2^n$ 
and induced matrix norm $||M||=\sqrt{\langle M,M\rangle}$ 
(as done in~\cite{boykin&roychowdhury:encryption}).
Note that $||M||=1$ if $M$ is unitary.
The set of all $\overline{\sigma_x}$ forms an orthonormal
basis for this vector space, so we get:
$$
p_i=||\sqrt{p_i}U_i||^2=
||\sum_x A_{ix}\frac{1}{\sqrt{2^{2n}}}\overline{\sigma_x}||^2
=\frac{1}{2^{2n}}\sum_x |A_{ix}|^2\leq \frac{1}{2^{2n}}.
$$
Hence $N\geq 2^{2n}$ and $\Hp{N}\geq 2n$.
\end{proof}

However, even granted this result it is still conceivable that a \PQC\
might require less randomness if it can ``spread out'' its encoding over 
many ancilla qubits --- it is even conceivable that those ancilla qubits 
can be used to {\em establish} privately shared randomness using some variant 
of quantum key distribution. The general case with ancilla is not addressed
in~\cite{boykin&roychowdhury:encryption}, and proving that the $2n$-bit lower
bound extends to this case requires more work. 
The next few theorems will do this.
These show that a \PQC\ that can transmit any $n$-qubit state 
requires $2n$ bits of randomness, no matter how many ancilla qubits it uses.
Thus Theorem~\ref{sufficient} exhibits an optimal quantum one-time pad,
analogous to the optimal classical one-time pad mentioned in the introduction.

We will use the notation $\C_k=\{\ket{i} \mid 0\leq i \leq k-1\}$
for the set of the first $k$ classical states.
The next theorem states that a \PQC\ that privately conveys 
$n$ qubits using $m$ bits of key, can be transformed into a 
\PQC\ that privately conveys any state from $\C_{2^{2n}}$, 
still using only $m$ bits of key.

\begin{theorem}\label{q2c}
If there exists a \PQC\ 
$[\H_{2^n},\E=\{ \sqrt{p_i} U_i \mid  1\leq i \leq N \},\ra,\rn]$, 
then there exists a \PQC\ 
$[\C_{2^{2n}},\E'=\{ \sqrt{p_i} U'_i \mid  1\leq i \leq N \},\ra,\tilde{I}_{2^n} \otimes \rn ]$.
\end{theorem}

\begin{proof}
For ease of notation we assume without loss of generality that
$\E$ uses no ancilla, so we assume $\rn$ is an $n$-qubit state and omit $\ra$
(this does not affect the proof in any way).
We first show that $\E(\ketbra{x}{y})=0$ whenever $x,y\in\C_{2^{2n}}$ and
$x\neq y$ ($\E(\ketbra{x}{y})$ is well-defined but somewhat of an abuse of 
notation, since the matrix $\ketbra{x}{y}$ is not a density matrix). 
This is implied by the following 3 equalities:\\[3mm]
$
\displaystyle
\rn = \E\left(\frac{1}{2}(\ketbra{x}{x}+\ketbra{y}{y})\right)  
    = \frac{1}{2}\left(\E(\ketbra{x}{x})+\E(\ketbra{y}{y})\right).\\[2mm]
\rn = \E\left((\frac{1}{\sqrt{2}}(\ket{x}+\ket{y}))(\frac{1}{\sqrt{2}}(\bra{x}+\bra{y}))\right)
= \frac{1}{2}\left(\E(\ketbra{x}{x})+\E(\ketbra{y}{y})+\E(\ketbra{x}{y}) +
\E(\ketbra{y}{x})\right).\\[2mm]
\rn = \E\left((\frac{1}{\sqrt{2}}(\ket{x}+i\ket{y}))(\frac{1}{\sqrt{2}}(\bra{x}-i\bra{y}))\right)
= \frac{1}{2}\left(\E(\ketbra{x}{x})+\E(\ketbra{y}{y})-i\E(\ketbra{x}{y})+i\E(\ketbra{y}{x})\right).\\[2mm]
$
The first and second equality imply $\E(\ketbra{x}{y}) + \E(\ketbra{y}{x})=0$, 
the first and third equality imply $\E(\ketbra{x}{y}) - \E(\ketbra{y}{x})=0$.
Hence $\E(\ketbra{x}{y})=\E(\ketbra{y}{x})=0$.

We now define $\E'$ and show that it is a \PQC.
Intuitively, $\E'$ will map every state from $\C_{2^{2n}}$ to a tensor product of 
$n$ Bell states by mapping pairs of bits to one of the four Bell states.%
\footnote{The 4 Bells states are $\frac{1}{\sqrt{2}}(\ket{00}\pm\ket{11})$ 
and $\frac{1}{\sqrt{2}}(\ket{01}\pm\ket{10})$.}
The second bits of the pairs are then moved to the second half
of the state and randomized by applying $\E$ to them.
Because of the entanglement between the two halves of each Bell state,
the resulting $2n$-qubit density matrix will be $\tilde{I}_{2^n} \otimes \rn$.
More specifically, define
$$
U \ket{x} = \left(\overline{\sigma_x} \otimes I_{2^n}\right)
\frac{1}{\sqrt{2^n}} \sum_{i=0}^{2^n-1}\ket{i}\ket{i},
$$
with $\overline{\sigma_x} = \sigma_{x_1} \otimes \cdots \otimes \sigma_{x_n}$ 
as in Theorem~\ref{sufficient}. Also define $U'_i=(I_{2^n} \otimes U_i)U$.
It remains to show that $\E'(\ketbra{x}{x})=\tilde{I}_{2^n} \otimes \rn$ 
for all $\ket{x}\in\C_{2^{2n}}$:
\begin{eqnarray*}
& & \E'(\ketbra{x}{x}) \\
&=& \sum_{i=1}^N p_i (I_{2^n} \otimes U_i) \left[
(\overline{\sigma_x} \otimes I_{2^n}) 
\left( \frac{1}{\sqrt{2^n}}\sum_{y=0}^{2^n-1} \ket{y}\ket{y} \right)
\left(  \frac{1}{\sqrt{2^n}}\sum_{z=0}^{2^n-1} \bra{z}\bra{z} \right)
(\overline{\sigma_x} \otimes I_{2^n})^{\dagger}\right] (I_{2^n} \otimes U_i)^{\dagger}\\
&=& (\overline{\sigma_x} \otimes I_{2^n}) \left[ \frac{1}{2^n} \sum_{i=1}^N p_i (I_{2^n} \otimes U_i)  
\left( \sum_{y,z\in\{0,2^n-1\}} \ketbra{y}{z} \otimes \ketbra{y}{z}  \right)
(I_{2^n} \otimes U_i)^{\dagger}\right] (\overline{\sigma_x} \otimes I_{2^n})^{\dagger}\\
&=& (\overline{\sigma_x} \otimes I_{2^n}) \left[   
\frac{1}{2^n}\sum_{y,z\in\{0,2^n-1\}} \ketbra{y}{z} \otimes 
\left( \sum_{i=1}^N p_i U_i \ketbra{y}{z} U_i^{\dagger} \right)
\right] (\overline{\sigma_x} \otimes I_{2^n})^{\dagger}\\ 
&=& (\overline{\sigma_x} \otimes I_{2^n}) \left[   
\frac{1}{2^n}\sum_{y,z\in\{0,2^n-1\}} \ketbra{y}{z} \otimes \E(\ketbra{y}{z})
\right] (\overline{\sigma_x} \otimes I_{2^n})^{\dagger}  \\ 
&\stackrel{(*)}{=} & (\overline{\sigma_x} \otimes I_{2^n}) \left[   
\frac{1}{2^n}\sum_{y=0}^{2^n-1} \ketbra{y}{y} \otimes \E(\ketbra{y}{y})
\right] (\overline{\sigma_x} \otimes I_{2^n})^{\dagger}  \\ 
&=& (\overline{\sigma_x} \otimes I_{2^n}) \left[   
\tilde{I}_{2^n} \otimes \rn
\right] (\overline{\sigma_x} \otimes I_{2^n})^{\dagger}\\
&=& \tilde{I}_{2^n} \otimes \rn.
\end{eqnarray*}
In the step marked by $(*)$ we used that $\E(\ketbra{y}{z})=0$ if $y\neq z$.
\end{proof}

Before proving a lower bound on the entropy required for sending arbitrary
$n$-qubit states, we first prove a lower bound on the entropy required
for sending states from $\C_{2^m}$. 
Privately sending any state from $\C_{2^m}$ corresponds to privately sending
any classical $m$-bit string.
If communication takes place through {\em classical} channels, then Shannon's 
theorem implies that $m$ bits of shared key are required to achieve such security.
Shannon's classical lower bound does not translate automatically to the quantum world
(it is in fact violated if a {\em two}-way quantum channel is available, 
see Footnote~\ref{noteQKD}). 
Nevertheless, if Alice and Bob communicate via a one-way quantum channel, 
then Shannon's theorem does generalize to the quantum world: 

\begin{theorem}\label{classical}
If $[\C_{2^m},\{ \sqrt{p_i} U_i \mid  1\leq i \leq N \},\ra,\rn]$ 
is a \PQC, then $\Hp{N} \geq m$.
\end{theorem}

\begin{proof}
Diagonalize the ancilla as $\ra=\sum_{j=1}^r q_j\ketbra{\psi_j}{\psi_j}$, 
so $S(\ra)=\Hq{r}$.
First note that the properties of Von Neumann entropy (Section~\ref{vne}) imply:
\begin{eqnarray*}
S(\rn) & = & S\left(\sum_{i=1}^N p_i U_i (\ketbra{0}{0} \otimes \ra)
U_i^{\dagger}\right) = S\left(\sum_{i=1}^N\sum_{j=1}^r p_iq_j
U_i (\ketbra{0}{0}\otimes\ketbra{\psi_j}{\psi_j}) U_i^{\dagger}\right)\\
       & \leq & H(p_1q_1,p_1q_2,\ldots,p_Nq_{r-1},p_Nq_r) = \Hp{N} + \Hq{r}.
\end{eqnarray*}
Secondly, note that
$$
S(\rn)  = S\left( \sum_{i=1}^N p_i  U_i (\tilde{I}_{2^m} \otimes \ra) U_i^{\dagger}\right) \geq \sum_{i=1}^N p_i S\left( \tilde{I}_{2^m} \otimes \ra\right) 
=\sum_{i=1}^N p_i (m+S(\ra)) = m+S(\ra).
$$
Combining these two inequalities gives the theorem.
\end{proof}

In particular, for sending arbitrary states from $\C_{2^{2n}}$ we need entropy
at least $2n$. Combining Theorems~\ref{q2c} and~\ref{classical} we thus obtain:

\begin{corollary}\label{main}
If $[\H_{2^n},\{ \sqrt{p_i} U_i \mid  1\leq i \leq N \},\ra,\rn]$ 
is a \PQC, then $\Hp{N}\geq 2n$ (and hence in particular $N\geq 2^{2n}$).
\end{corollary}

In relation to Theorem~\ref{reel}, note that $\C_{2^n}\subseteq B^{\otimes n}$.
Hence another corollary of Theorem~\ref{classical} is the optimality
of the \PQC\ of Theorem~\ref{reel}:

\begin{corollary}\label{reel_lower}
If $[B^{\otimes n},\{ \sqrt{p_i} U_i \mid  1\leq i \leq N \},\ra,\rn]$ 
is a \PQC, then $\Hp{N}\geq n$ (and hence in particular $N\geq 2^{n}$).
\end{corollary}

\section{Randomization of Quantum States}\label{rnd}

The above concepts and results were motivated by cryptographic goals, namely 
to enable private transmission of quantum information using a shared classical key.
However, our results can also be stated in terms of the problem of 
``forgetting'' or ``randomizing'' quantum information, 
as discussed recently by Braunstein, Lo, and Spiller~\cite{bls:forget}.

The {\em randomization} of a quantum source $\S$ is a procedure that 
maps any state $\rho$ coming from $\S$ to some fixed constant 
state $\rho_0$ (for instance the completely mixed state).
The process thus ``forgets'' what was specific to $\rho$.
To help in this process, we allow the randomizing process to make use of 
a piece of the environment which is in some fixed state $\ra$ (ancilla qubits).
We also give it access to some source of classical randomness.
Because every quantum operation can be viewed as a unitary transformation
on a larger space, we can assume without loss of generality that 
the randomization process has the following form:
it uses the source of randomness to pick some $i$ with probability
$p_i$, then it applies some unitary transformation $U_i$ to $\rho$
and the ancillary environment, and then it forgets $i$.
The resulting mixed state should be $\rho_0$.
At this point it should be clear to the reader that if 
$[\S,\{ \sqrt{p_i} U_i \mid  1\leq i \leq N \},\ra,\rn]$ is a \PQC, 
then it also constitutes a randomization procedure, and vice versa.

We are interested in the amount of entropy that such a randomization procedure 
needs to generate. This is the entropy of forgetting the random classical input $i$.
It quantifies the thermodynamic cost of the process. 
Braunstein, Lo, and Spiller~\cite{bls:forget} have shown that 2 bits 
of entropy are necessary and sufficient for the randomization of 1 qubit.
By translating our \PQC-results to the randomization-context, 
we can generalize their result to:

\begin{corollary}
The generation of $2n$ bits of entropy is sufficient and necessary in order 
to randomize arbitrary $n$-qubit states.
\end{corollary}

\begin{proof}
Sufficiency follows from Theorem~\ref{sufficient} and necessity from Corollary~\ref{main}. 
\end{proof}

For the more limited set of states $\S=B^{\otimes n}$ we have:

\begin{corollary}
The generation of $n$ bits of entropy is sufficient and necessary in order 
to randomize arbitrary tensor products of $n$ real-amplitude qubits.
\end{corollary}

\begin{proof}
Sufficiency follows from Theorem~\ref{reel} and necessity from Corollary~\ref{reel_lower}. 
\end{proof}

\section{Summary}

The main result of this paper is an optimal quantum version of the classical
one-time pad. On the one hand, if Alice and Bob share $2n$ bits of key, Alice can
send Bob any $n$-qubit state $\rho$, encoded in another $n$-qubit state in a
way which conveys no information about $\rho$ to the eavesdropper.
This is a simple scheme which works locally (i.e.~deals with each qubit separately)
and uses no ancillary qubits.
On the other hand, we showed that even if Alice and Bob are allowed to use
any number of ancilla qubits, then they still require $2n$ bits of entropy.
In the context of state randomization, it follows that $2n$ bits of entropy
are necessary and sufficient for randomization of $n$-qubit states.

\subsection*{Acknowledgment}
We thank Richard Cleve, Hoi-Kwong Lo, Michael Nielsen, Harry Buhrman,
and P.~Oscar Boykin for useful discussions and comments.


\newcommand{\etalchar}[1]{$^{#1}$}

\end{document}